\documentclass[journal=jacsat,manuscript=article]{achemso}
\usepackage{pdfpages}
\UseRawInputEncoding

\usepackage{braket}
\usepackage{bbold}
\usepackage{soul}
\usepackage{color}
\usepackage{amsmath}
\usepackage{latexsym}
\usepackage{amssymb}
\usepackage{graphics,epstopdf}
\usepackage{hyperref}
\usepackage{xcolor}
\usepackage[version=3]{mhchem} 
\hypersetup{hidelinks}

\author{Song Chen}
\affiliation{School of Physics and Wuhan National High Magnetic Field Center,
Huazhong University of Science and Technology, Wuhan 430074, People's Republic of China.}

\author{Hua-Hua Fu}
\email{hhfu@hust.edu.cn}
\affiliation{School of Physics and Wuhan National High Magnetic Field Center,
Huazhong University of Science and Technology, Wuhan 430074, People's Republic of China.}
\alsoaffiliation{Institute for Quantum Science and Engineering,
 Huazhong University of Science and Technology, Wuhan, Hubei 430074, China.}

\title[An \textsf{achemso} demo]
{Chirality-dependent persistent spin current in single circular helix molecules}

\makeatletter
\newcommand*{\forcekeywords}{
  \acs@keywords@print
  \let\acs@keywords@print\relax
}
\makeatother

\begin{document}




\begin{abstract}
  \textbf{Since the chiral-induced spin selectivity (CISS) was first observed experimentally, its microscopic mechanism has been continuously explored by the scientific community. Among these investigations, the non-equilibrium effects and the unknown origins of spin-orbit coupling (SOC) have been the central issues discussed in recent years. Here, we have achieved a persistent spin current in different circular single-helix molecule driven by a magnetic field, which is inherently linked to the equilibrium state associated with chirality. Due to its measurement method being different from the transport currents observed in previous experiments, the origin of its spin-orbit coupling can be explored by modifying the substrate with different light and heavy metal elements. Our results demonstrate that a persistent spin current can be observed regardless of whether the SOC originates from the chiral molecule or the substrate. Furthermore, by tuning the direction of the magnetic field, we can achieve a phase transition between trivial and non-trivial chiral persistent spin currents. Our work provides a new perspective and platform for exploring the nature of CISS and controlling the effects of CISS.}
\end{abstract}

\section{Introduction}
The subtle relationship between spin selectivity and chirality has been appreciated by researchers ever since 1999, when Naaman and collaborators observed it in self-organized monolayers of chiral 
molecules, giving rise to the chirality-induced spin selectivity (CISS) effect, wherein specific spin orientations of electrons emerge as spin-degenerate electrons passing through chiral molecules.\cite{Ray1999} This phenomenon is typically associated with the structural chirality and spin-orbit coupling (SOC). Due to CISS intriguing nature and potential applications in molecular spintronics, it has been continuously explored by various theories \cite{1,2,3,4,5,6,7,7.1,8,8.1, 9,10,10.1, 11, 12,13,14,15,ChenFu} and experiments. \cite{Ray1999, 1.1, 1.2, 1.3,1.4,1.5,1.6,1.7,1.8,1.9,1.10,1.11,1.12,1.13,1.14} Nonetheless, the underlying micro-mechanisms of this phenomenon remain unclear at present.

In recent years, two key characteristics within the CISS phenomenon, holding promise for understanding its micro-mechanisms, have gained attention in various scholarly studies. First, 
the non-equilibrium phenomena manifested in the current CISS experiments, including various types of measurement methods \cite{1.4, 1.5, 1.7, 1.10,1.11,1.12,1.13,1.14} and electron transfer under external driving forces, such as light-illumination, \cite{Ray1999, 1.1, 1.2, 1.3} voltage drop \cite{1.7,1.8,1.9}  and substrate contact.\cite{1.6}
Second, the question arises whether the source of the SOC stems from the chiral molecule or the substrate. $(i)$ In earlier theoretical explorations, SOC was considered as an intrachain product induced by chirality. But given that the SOC strength of carbon atom p-electrons is only about 6 meV, which is insufficient to explain the strong spin polarization observed experimentally, numerous theoretical endeavors have explored enhancements via mechanisms like electron-phonon interactions\cite{7,7.1}, electron-electron correlations\cite{8}, polarons\cite{9}, frictional dissipation\cite{10}, spin Fano resonance\cite{10.1} and Berry force induced by coupled nuclear-spin dynamics near a conical intersection\cite{11}, among others. Additionally, Li $et$ $al.$ present a strong SOC arises from electron-hole pairing driven by many-body correlation, potentially accounting for the observed spin polarization at room temperature.\cite{12} $(ii)$ In the context of early experimental endeavors, the protein bacteriorhodopsin, which is comprised of seven $\alpha$-helices oriented perpendicularly to the membrane surface, was employed for photoelectron experiments utilizing aluminum covered by its natural oxide on the surface as the substrate, continuing to reveal a pronounced spin polarization phenomenon.\cite{1.2} Consequently, the focus has persistently shifted towards chiral molecules in the quest to unveil the SOC origins. Until recently, Liu $et$ $al.$ noted a distinction: in photoelectron experiments, Mott detectors cannot differentiate between spin and orbital angular momentum, resulting in a combined magnetization, which enables even light metal substrates to display substantial polarization phenomena. But transport experiments exclusively measure spin polarization, necessitating heavy metal substrates \cite{13}.

On the flip side, it should be emphasized that nearly all theoretical models previously used to investigate the CISS effect have been based on straight linear helical chains, potentially leading us to overlook the discovery of other intriguing phenomena associated with or induced by CISS that could have been observed within circular helical molecular systems, such as the spin destructive quantum interference effect \cite{ChenFu}. Additionally, circular DNA and proteins have been widely observed in nature.\cite{Harrison, Tischer, Shulman, Martnez, David} With the  development of nanotechnology, many structured proteins and DNA can be synthesized experimentally \cite{Doyle, Derrick}.

The persistent current (PC) in the many-body ground state of a quantum ring with enclosed magnetic flux represents an equilibrium current in the absence of external excitation. They are detectable solely when the coherence length of the electrons surpasses the ring's circumference. The current's orientation alternates based on the strength of the field, exhibiting periodicity with a flux quantum. 
Furthermore, as electrons traverse a ring with SOC, the electron's spin acquires a spin Berry phase, resulting in a persistent spin current (PSC). 
While early theoretical \cite{von,cheung,ambegaokar,cedraschi,schutz,schmeltzer, sun2007,sun2008,tokatly} and experimental \cite{jariwala,levy,mailly,kleemans,bluhm,bleszynski,castellanos} research primarily concentrated on mesoscopic metal rings, the recent appearance of PC in ultracold atomic gases \cite{Ryu,Beattie,cai,pactu,del} and, more significantly, in aromatic nanomolecules \cite{peeks,kopp,rickhaus,asademont,ren} has initiated a fresh cycle of theoretical research.
Peeks $et$ $al.$ demonstrate a six-porphyrin nanoring's antiaromatic to aromatic shift, showcasing global ring current induction via oxidation manipulation and revealing room-temperature ring currents as quantum coherence evidence in large molecules \cite{peeks}.
Judd $et$ $al.$ employed STM and STS on a silver surface to characterize the electronic structure of a 40-porphyrin nanoring, revealing coherent electron delocalization and suggesting its potential as a molecular quantum ring \cite{judd}.

\begin{figure}
\includegraphics[width=11cm]{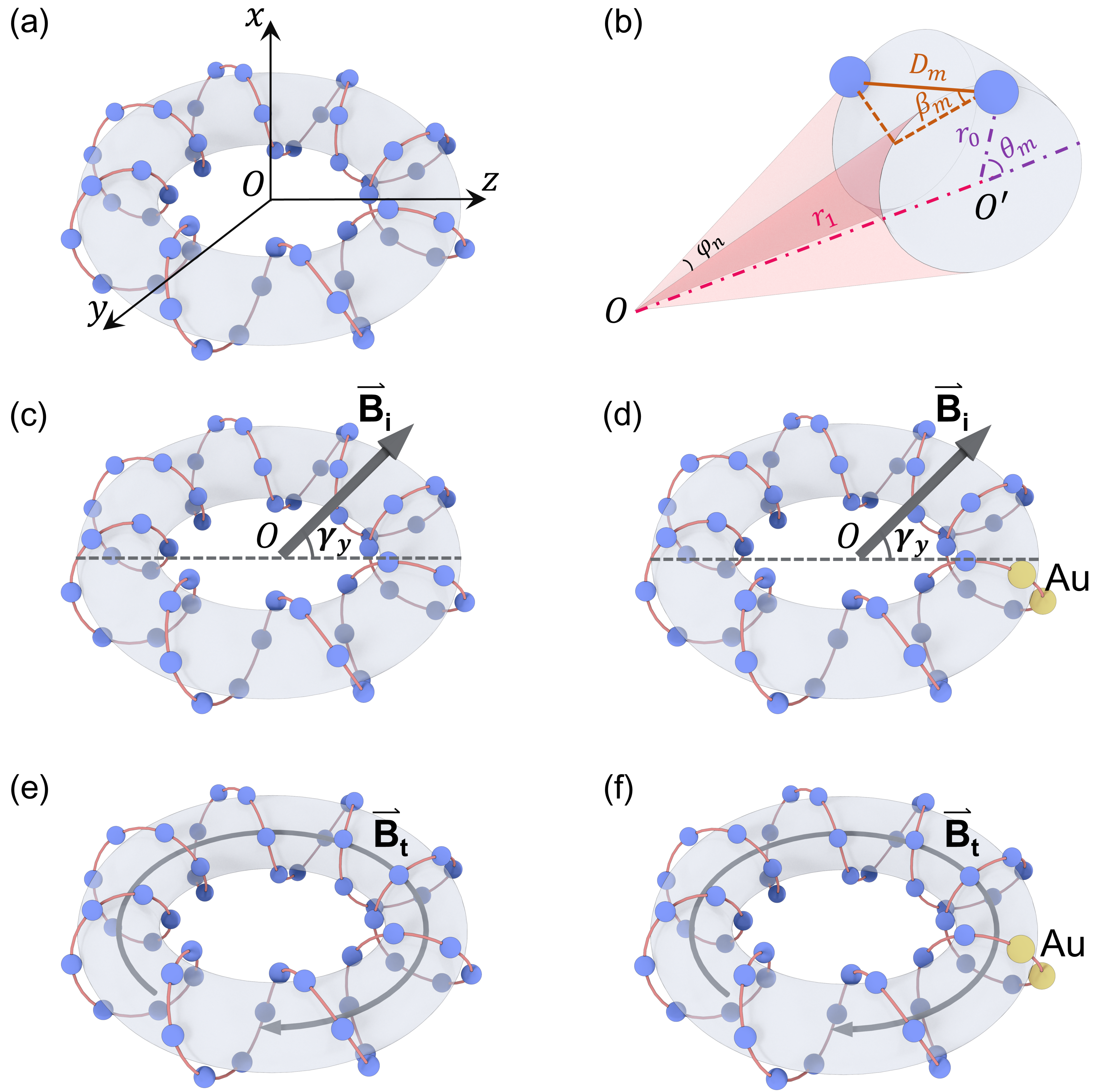}
\caption{\textcolor{black}{Theoretical model of CISS effect in the CHMs under various magnetic field shapes. (a) Side schematic view of a single CHM. (b) Geometric magnification diagram of the connections between nearest neighboring lattice point and the central $O$ point on the CHM in (a). Here, $O$ ($O^{\prime}$) and $r_1$ ($r_0$) denote the center and radius of the toridal chiral molecule and of its crossing-section plane. $\phi_n$ and $\theta_m$ indicate the toroidal and poloidal angle. $\beta_m$ and $D_m$ represent the space angle and the distance between two atoms located in the above plane. Under an inclined magnetic field $\vec{\textbf{B}}_i$, the chain-internal chiral-induced SOC (c) and the SOC provided by the heavy metal Au substrate (d) are illustrated in the diagram. (e) and (f) represent similar theoretical models, but under a toroidal  magnetic field $\vec{\textbf{B}}_t$.}}
\label{fig1}
\end{figure}

In this work, we establish two theoretical models under different magnetic field shape, along with their corresponding effective tight-binding Hamiltonians, taking into consideration chiral-induced SOC and substrate-provided SOC, to study CISS effect in various CHMs without relying on electrodes, as described in Figure~\ref{fig1}(a). 
By employing the Green's function techniques,  we investigated persistent charge current (PCC) and chirality-related PSC, revealing that in chiral-induced SOC models, it is necessary to satisfy at least one of the conditions, $\mathcal{N} \bmod M \neq 0$, asymmetric magnetic field, or long-range hopping, to generate chirality-related PSC within the angular range of $[0,\pi/2) \cup (\pi/2,\pi]$ relative to the molecular plane, while substrate-induced SOC models allow the generation of both PCC and chirality-related PSC without specific conditions.
More importantly, when the direction of the magnetic field varies within the range $[0, \pi]$, both models exhibit the same behavior in transitioning between chirality-related PSC and chirality-independent PSC. Finally, we extensively discuss various experimental-related issues.

\section{Results and discussion}

\subsection{Intrachain SOC in CSH under inclined magnetic field}

We consider a closed CSH under a magnetic field at an angle of $\gamma_y (\gamma_y \ne 0)$ with respect to the molecular plane , as drawn in Figure~\ref{fig1} c.
The Hamiltonian for the electrons in the presence of chirality-induced SOC and a magnetic field is gievn by $\mathcal{H}=\mathcal{H}_{c}+\mathcal{H}_{\mathrm{soc}}$. Here, we disregard the Zeeman splitting induced by an external field, which we will address later in the final discussion. 

The first term $\mathcal{H}_{c}=\frac{1}{2m_e}(\mathbf{p}-e\mathbf{A})^2 + V$, presents an account of the electrons' kinetic and potential energies within the context of CSH, where $\mathbf{p}$ denotes the momentum operator, and $\mathbf{A}$ represents the vector potential. By using the second quantization, $\mathcal{H}_{c}$ is discretized as 
\begin{equation}
\mathcal{H} _c=\sum_{n=1}^{\mathcal{N}}{\varepsilon _nc_{n}^{\dagger}c_n}+\sum_{n=1}^{\mathcal{N}}{\sum_j^J{\left( t_{nj}c_{n+j}^{\dagger}e^{i\varTheta_j}c_n+\mathrm{H}.\mathrm{c}. \right)}}
\end{equation}
 where $c_{n}^{\dagger}=(c_{n \uparrow}^{\dagger}, c_{n \downarrow}^{\dagger})$ and $\varepsilon_n$ are the creation operator and on-site energy for the electrons at the $n$-th lattice in CSH whose length is $\mathcal{N}$, with $c_{1}^{\dagger} = c_{\mathcal{N}+1}^{\dagger}$. $t_{nj}$ is the $j$th neighboring hopping integral at the $n$-th lattice site and owing to the exponential decay of the wave function, $t_{nj}$ which is defined by the spatial distance $d_{nj}$ amid the $n$th and $(n+j)$th lattice sites, can be written as $t_{nj} = t_{n}$exp$(-\alpha d_{nj})$, where the coefficients $\alpha$ are determined by an individual atomic Wannier function.
 The phase factors arising from the Aharonov-Bohm (AB) flux phase, are expressed as $\varTheta_j = 2\pi j\phi/\mathcal{N}\phi_0$, where flux quantum $\phi_0 = h/e$. Additionally, the magnetic flux $\phi$ in this case is contributed by both the loop encircled by the helical axis and the regions between the $n$th and $(n+j)$th sublattices of CSH.

The second component, $\mathcal{H}_{\mathrm{soc}}=\frac{\hbar}{4 m_e^2 c^2} \nabla V \cdot\hat{\sigma} \times (\mathbf{p}-e\mathbf{A})$, represents the chirality-induced SOC. Here, $c$ is the speed of light and $\hat{\sigma} = (\sigma_x, \sigma_y, \sigma_z)$  corresponds to the Pauli matrices. An effective spin quantization axis determined by the external magnetic field's orientation combined with the strength and gradients of the SOC. But in this study, we primarily focus on establishing the spin quantization axis based on the dominating impact of the magnetic field, owing to the potential to examine magnetic fields of magnitudes significantly surpassing the strength of SOC. the feasibility of such an approach will be elaborated upon in detail during the concluding discussion. Parallel to the aforementioned, $\mathcal{H}_{\mathrm{soc}}$ can be discretized based on our previous model as follows \cite{ChenFu}
\begin{equation}
\mathcal{H}_{\mathrm{soc}}=\sum_{n=1}^{\mathcal{N}}{\sum_j^J}\sum_{p}^{\left\{ 1,-1 \right\} }{2i\lambda_{nj} \cos \left( \theta _{m}^{-} \right) \cos \left( \varphi _{n}^{-} \right)  c_{n+j}^{\dagger}e^{i\varTheta_j}H[p(n-\frac{\mathcal{N}}{2})]\sigma _{n,j,p}^mc_n}+\mathrm{H}.\mathrm{c}.
\end{equation}
where $\lambda_{nj} = \lambda_{n}$exp$(-\alpha d_{nj})$; $H[x]$ represents the Heaviside step function and $\sigma_{n,j,p}^m(\theta, \varphi)=(\sin \gamma_y \sin \theta_{m,j}^{+} \sin \varphi_{n,j}^{+} \sin \beta_{m,j} + \cos \gamma_y \cos \theta_{m,j}^{+} \sin \beta_{m,j} + \sin \gamma_y \cos \varphi_{n,j}^{+} \cos \beta_{m,j}) \sigma_x + [\cos \gamma_x (\sin \varphi_{n,j}^{+} \\ \cos \beta_{m,j}-\sin \theta_{m,j}^{+} \cos \varphi_{n,j}^{+} \sin \beta_{m,j}) + \sin p\gamma_x (\sin \gamma_y \cos \theta_{m,j}^{+} \sin \beta_{m,j} - \cos \gamma_y \sin \theta_{m,j}^{+} \sin \varphi_{n,j}^{+} \sin \beta_{m,j}- \cos \gamma_y \cos \varphi_{n,j}^{+} \cos \beta_{m,j})] \sigma_y + [-\sin p\gamma_x (\sin \varphi_{n,j}^{+} \cos \beta_{m,j}-\sin \theta_{m,j}^{+} \cos \varphi_{n,j}^{+} \sin \beta_{m,j}) + \cos \gamma_x (\sin \gamma_y \\\cos \theta_{m,j}^{+} \sin \beta_{m,j} - \cos \gamma_y \sin \theta_{m,j}^{+} \sin \varphi_{n,j}^{+} \sin \beta_{m,j}- \cos \gamma_y \cos \varphi_{n,j}^{+} \cos \beta_{m,j})] \sigma_z$, here $\gamma_x$ is the angle of rotation of the magnetic field around the x-axis; $\varphi _{n,j}^{\pm}=( \varphi _{n+j} \pm \varphi _n ) /2; \varphi _n=\left( n-1 \right) \Delta \varphi ; \Delta \varphi =2\pi /\mathcal{N}$ and $\theta _{m,j}^{\pm}=\left( \theta _{m+j}\pm \theta _m \right) /2; \theta _m=\left( m-1 \right) \Delta \theta ; \Delta \theta =2\pi /M$, $M$ the number of atoms in every unit cell. 

To provide specificity, we first consider a circular single-stranded DNA (cssDNA) molecule with model parameters $M=10$, $r_0 = 7$$\textup{~\AA}$, $r_1 =\mathcal{N}h/2\pi$, here $h=3.32$$\textup{~\AA}$. Moreover, $\varepsilon_n = 0$ and the hopping parameter $t_{n}$ corresponding to the minimum distance within a unit cell is considered as $t$, and thus the SOC is estimated to be $\lambda_{soc}$=$0.1t$. For B-form DNA, we consider a decay exponent of $\alpha = 1.11$. Some other structural parameters are described in Supplementary Information (SI), Sec. B.

\begin{figure}[t]
\includegraphics[width=16.5cm]{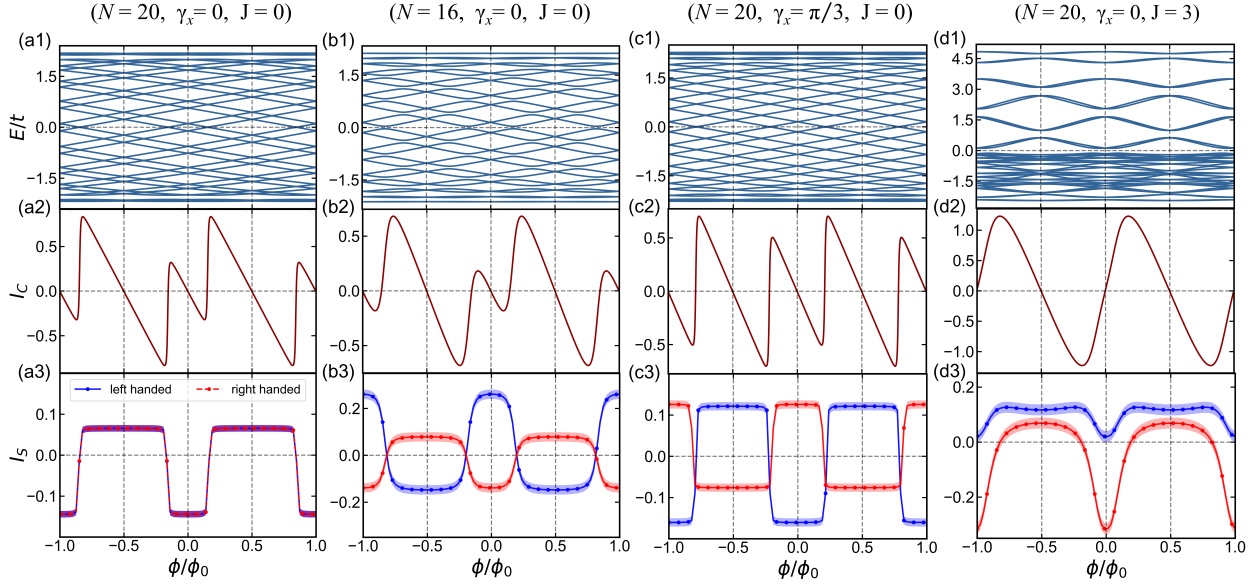}
\caption{\textcolor{black}{In the model of chain-internal chiral-induced SOC, the energy spectrum, persistent charge current, and persistent spin current of different chirality vary under different conditions as the magnetic flux strength changes with $\gamma_y = \pi/6$ in an inclined magnetic field. For $N=20$, $\gamma_x = 0$, and $J = 0$, corresponding to (a1)-(a3), the remaining cases involve changing only one of these three conditions at a time to investigate the variations in these physical quantities, $N=16$((b1)-(b3)), $\gamma_x = \pi/3$((c1)-(c3)), $J = 3$((d1)-(d3)). The model parameters are $M=10$, $r_0 = 7$$\textup{~\AA}$, $r_1 =\mathcal{N}h/2\pi$, $\lambda_{soc}$=$0.1t$ and $\alpha = 1.11$.} }
\label{fig2}
\end{figure}
We investigat the occurrence of PC in cssDNA under a magnetic field when $\gamma_y = \pi/6$, as described in Figure~\ref {fig1}(c). Figure~\ref{fig2} illustrates the behavior of spectrum, PCC and PSC with varying magnetic flux $\phi$, under the conditions of different $\mathcal{N}$, $\gamma_x$ and the presence or absence of long-range hopping (LRH). The gauge invariance ensures  that they all exhibit characteristic periodic variations with respect to $\phi$, with one flux quantum $\phi_0 = h/e$ as the period. The generated ring charge current is directly related to the gradient of the overall ground state energy (Figures~\ref{fig2}(a1)-(d1)) of the system concerning the magnetic field, $I_{c} =-\sum{\frac{\partial \mathrm{E}}{\partial \mathrm{\phi}}}$, and this current oscillates between diatropic (negative) and paratropic (positive) as the field strength increases. Additionally, under time invariance, it is easy to have $I_c(-\phi)=-I_c(\phi)$. It's noteworthy that PCC exhibits significant distinctions between the absence (Figures~\ref{fig2}(d2) and presence (Figures~\ref{fig2}(a2)-(c2)) of LRH due to its reliance on the number of energy levels below the Fermi energy and the slope according to the definition provided above. The renormalization of the hopping probabilities that occurs when multiple hopping are active takes place non-uniform for each state and breaks the electron-hole symmetry, thereby causing the energy level distribution to lose symmetry with respect to the Fermi level. As a result, this leads to the generation of distinct maximum and minimum values for PCC that are noticeably different from the others.
PSC exists when inversion symmetry is broken due to chirality, much like how time-reversal symmetry breaking leads to the PCC. Under the condition of $\mathcal{N} \bmod M=0$ (i.e., $\mathcal{N} = 20$), all electrons' spin in the molecule experiencing a magnetic field in the same direction ($\gamma_x = 0$), and absence LRH ($J = 0$), the right-handed cssDNA and left-handed molecules, when subjected to transformations of the twist and spatial angle through reflection symmetry denoted as from $\theta$ to -$\theta$ and $\beta$ to $\pi$-$\beta$ respectively, exhibit the same PSC,  as described in Figures~\ref{fig2} (a3). 
However, when at least one of these three conditions is not met, the right-handed PSC and left-handed one exhibit noticeable distinctions, although they are not opposite to each other, as illustrated in Figures~\ref{fig2}(b3)-(d3).

\subsection{Substrate SOC in CSH under inclined magnetic field}

Another mainstream hypothesis regarding the origin of SOC supports that one arises from the heavy metal substrate or electrode. Here, we present a model based on this hypothesis, in which we consider assembling $3_{10}$ helix protein molecules on both sides of a gold substrate with the thiol bond, and placing them collectively in a magnetic field oriented at an angle of $\gamma_y$ with respect to the molecular plane, as drawn in Figure~\ref{fig1} d. The Hamiltonian of electrons under the influence of heavy metal element substrate SOC and a magnetic field is provided as follows


\begin{equation}
\begin{aligned}
\mathcal{H}& =\sum_{n=1}^{\mathcal{N}}\left[\varepsilon_{n, p}+\left(\lambda_{s u b} \delta_{0,\left\lfloor\frac{n}{3}\right\rfloor}+\lambda_{m o l} \delta_{1, \operatorname{sgn}(\left\lfloor\frac{n}{3}\right\rfloor)}\right) h_{s o c}\right] c_n^{\dagger} c_n \\
& +\sum_{n=1}^{\mathcal{N}}\left[\left(T_{n, p}^{s u b} \otimes \mathbb{1}_2\right) \delta_{0,(n-\mathcal{N})\left\lfloor\frac{n}{3}\right\rfloor}+\left(T_{n, p}^{m o l} \otimes \mathbb{1}_2\right) \delta_{1, \operatorname{sgn}(|n-\mathcal{N}|\left\lfloor\frac{n}{3}\right\rfloor)}\right] e^{i \varTheta} c_{n+1}^{\dagger} c_n+\text { H.c. } \\
&
\end{aligned}
\end{equation}
where $c_{n}^{\dagger}=(c_{n p_x\uparrow}^{\dagger}, c_{n p_x\downarrow}^{\dagger},c_{n p_y\uparrow}^{\dagger}, c_{n p_y\downarrow}^{\dagger},c_{n p_z\uparrow}^{\dagger}, c_{n p_z\downarrow}^{\dagger})$ and $\varepsilon_{np}$ are the creation operator and on-site energy for the distinct electron orbitals at the $n$-th site in CSH with a length of $\mathcal{N}$.  
$h_{s o c}$ and $\mathbb{1}_2$ respectively refer to the matrices of the SOC operator in the $p_{x,y,z}$ basis, and the  $2 \times 2$ identity matrix. $\left\lfloor x\right\rfloor$ corresponds to the floor function, while $\operatorname{sgn}(x)$  corresponds to the sign function. $\lambda_{s u b}$ and $\lambda_{m o l}$ represent the SOC strength of the $p$ orbitals for the substrate and chiral molecules, respectively. $T_{n, p}^{s u b}$ and $T_{n, p}^{mol}$ respectively denote the nearest-neighbor hopping matrices for the substrate and chiral molecules on different electron orbitals and can be expressed as
\begin{equation}
\begin{aligned}
T_{n,p,j}^{s/m}=\left( \begin{matrix}
	t_{nj,xx}^{s/m}&		t_{nj,xy}^{s/m}&		t_{nj,xz}^{s/m}\\
	t_{nj,yx}^{s/m}&		t_{nj,yy}^{s/m}&		t_{nj,yz}^{s/m}\\
	t_{nj,zx}^{s/m}&		t_{nj,zy}^{s/m}&		t_{nj,zz}^{s/m}\\
\end{matrix} \right) 
\end{aligned}
\end{equation}

\begin{equation}
\begin{aligned}
\begin{array}{ll}
t_{n j, x x}^{s / m}=l^2 V_{p p \sigma}^{s / m}+\left(1-l^2\right) V_{p p \pi}^{s / m} & t_{n j, x y}^{s / m}=t_{n j, y x}^{s / m}=\operatorname{lm}\left(V_{p p \sigma}^{s / m}-V_{p p \pi}^{s / m}\right) \\
t_{n j, y y}^{s / m}=m^2 V_{p p \sigma}^{s / m}+\left(1-m^2\right) V_{p p \pi}^{s / m} & t_{n j, y z}^{s / m}=t_{n j, z y}^{s / m}=m n\left(V_{p p \sigma}^{s / m}-V_{p p \pi}^{s / m}\right) \\
t_{n j, z z}^{s / m}=n^2 V_{p p \sigma}^{s / m}+\left(1-n^2\right) V_{p p \pi}^{s / m} & t_{n j, z x}^{s / m}=t_{n j, x z}^{s / m}=n l\left(V_{p p \sigma}^{s / m}-V_{p p \pi}^{s / m}\right)
\end{array}
\end{aligned}
\end{equation}
here the direction cosines of vectors from lattice site $n$ to $n+1$ are represented by $l$, $m$, and $n$. Position vectors is written $\vec{r}_n=(l \hat{x}+m \hat{y}+n \hat{z}) d=(-\sin \gamma_y\left(r_1+r_0 \cos \theta\right) \sin \varphi+\cos \gamma_y r_0 \sin \theta) \hat{x}+\left(r_1+r \cos \theta\right) \cos \varphi \hat{y}+\left(\cos \gamma_y\left(r_1+r_0 \cos \theta\right) \sin \varphi+\sin \gamma_y r_0 \sin \theta\right) \hat{z}$.

\begin{figure}[t]
\includegraphics[width=16cm]{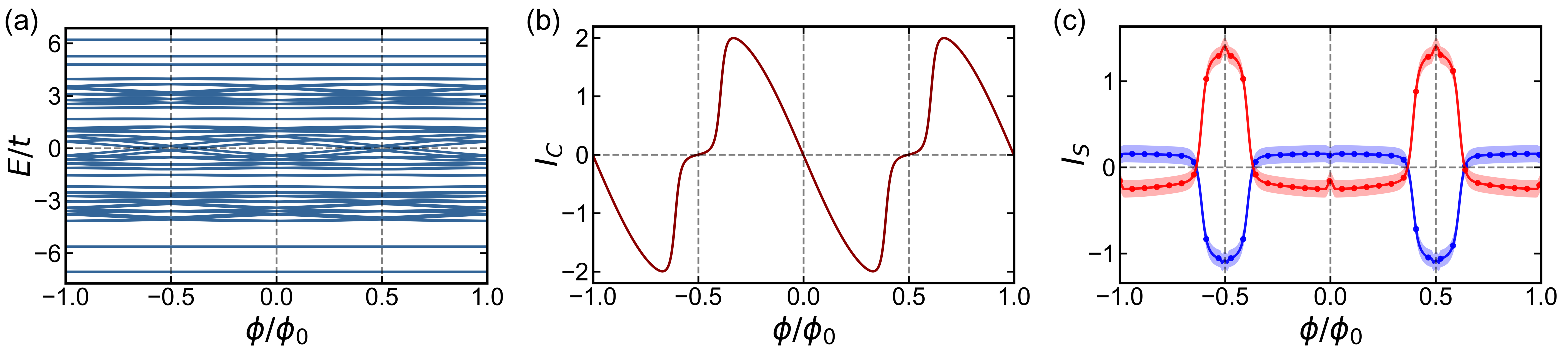}
\caption{\textcolor{black}{In the model of substrate-induced SOC, the energy spectrum, persistent charge current, and persistent spin current of different chirality vary under different conditions as the magnetic flux strength changes with $\gamma_y = \pi/6$ in an inclined magnetic field. The energy parameters used here are $V_{p p \sigma}^{m} = -3t$, $V_{p p \sigma}^{s} = -2t$ and $ V_{p p \pi}^{s} = 3t$, $\lambda_{s u b} = 2t$ and $\lambda_{m o l} = 0.012t$. The remaining geometric parameters are $M=3$, $r_0 = 1.9$$\textup{~\AA}$, $r_1 =\mathcal{N}h/2\pi$, here $h=2.0$$\textup{~\AA}$.}}
\label{fig3}
\end{figure}

Similar to the previous description, we express the hopping parameters $ V_{p p \pi}^{m}$ in units of energy, denoted as $t$ ($t\approx -0.5$ $\mathrm{eV}$). For other energy parameters, $\varepsilon_n$, $V_{p p \sigma}^{m}$, $V_{p p \sigma}^{s}$ and $ V_{p p \pi}^{s}$ are $0$, $-3t$, $-2t$ and $3t$ respectively. $\lambda_{s u b}$ and $\lambda_{m o l}$ are $2t$ and $0.012t$, respectively. The structural parameters are as follows $M=3$, $r_0 = 1.9$$\textup{~\AA}$, $r_1 =\mathcal{N}h/2\pi$, here $h=2.0$$\textup{~\AA}$. It is evident that the energy spectrum and current shown in Figure~\ref{fig3} share almost identical characteristics with those in Figure 2, except for numerical differences. Given the presence of gold atoms in the ring, which are distinct from the carbon atoms in chiral molecules, there is no need for us to engage in the classification discussions as done earlier. 
In this extreme case, when we consider all atoms in this model as identical with strong SOC, leading to the outcomes presented in Figure~\ref{fig2} involving a discussion of values for ($N$, $\gamma_x$, $J$), different considerations can result in either trivial or nontrivial chiral persistent spin currents. More importantly, with entirely different physical picture and construction processes, the two models generate similar physical outcomes, where the former involves chiral-induced SOC that electrons experience during tunneling within the molecular potential, and the latter incorporates chirality solely affecting electron hopping vectors.

\subsection{Intrachain SOC in CSH under toroidal magnetic field}

In what follows, considering a specific scenario in which the magnetic field is parallel ($\gamma_y = 0$) to the plane of the chiral molecule, we contemplate a toroidal magnetic field aligned parallel to the helical axis, as the magnetic flux contribution arises solely from within the molecule in this case. At this point, we modify our SOC Hamiltonian as follows


\begin{equation}
\mathcal{H}_{soc}^t = \sum_{n=1}^{\mathcal{N}} \sum_j^J 2 i \lambda_{nj} \cos \left(\theta_m^{-}\right) \cos \left(\varphi_n^{-}\right) \cos \left(\gamma_{x, n}^{-}\right) c_{n+j}^{\dagger} e^{i\varTheta_j} \sigma _{n,m,j}^t c_n c_n+\mathrm{H}.\mathrm{c}.
\end{equation}
where, $\sigma _{n,m,j}^t = \cos \theta _{m,j}^{+}\sin \beta _{m,j}\sigma _x-( \cos \gamma _{x,n}^{+}\sin \theta _{m,j}^{+}\sin \varphi _{n,j}^{+}\sin \beta _{m,j}+\cos \gamma _{x,n}^{+}\cos \varphi _{n,j}^{+}\cos \beta _{m,j}+\sin \gamma _{x,n}^{+}\sin \varphi _{n,j}^{+}\cos \beta _{m,j}-\sin \gamma _{x,n}^{+}\sin \theta _{m,j}^{+}\cos \varphi _{n,j}^{+}\sin \beta _{m,j} ) \sigma _y+( \sin \gamma _{x,n}^{+}\sin \theta _{m,j}^{+}\sin \varphi _{n,j}^{+}\sin \beta _{m,j}+\sin \gamma _{x,n}^{+}\cos \varphi _{n,j}^{+}\cos \beta _{m,j}+\cos \gamma _{x,n}^{+}\sin \theta _{m,j}^{+}\cos \varphi _{n,j}^{+}\sin \beta _{m,j}-\cos \gamma _{x,n}^{+}\sin \varphi _{n,j}^{+}\cos \beta _{m,j} ) \sigma _z.
$
here, $\gamma _{x,n}^{\pm}=\left( \gamma _{n+1}\pm \gamma _n \right) /2; \gamma _{x,n}\left( n-1 \right) \Delta \varphi.$ The remaining variables are consistent with the equation (2).

\begin{figure}[t]
\includegraphics[width=15cm]{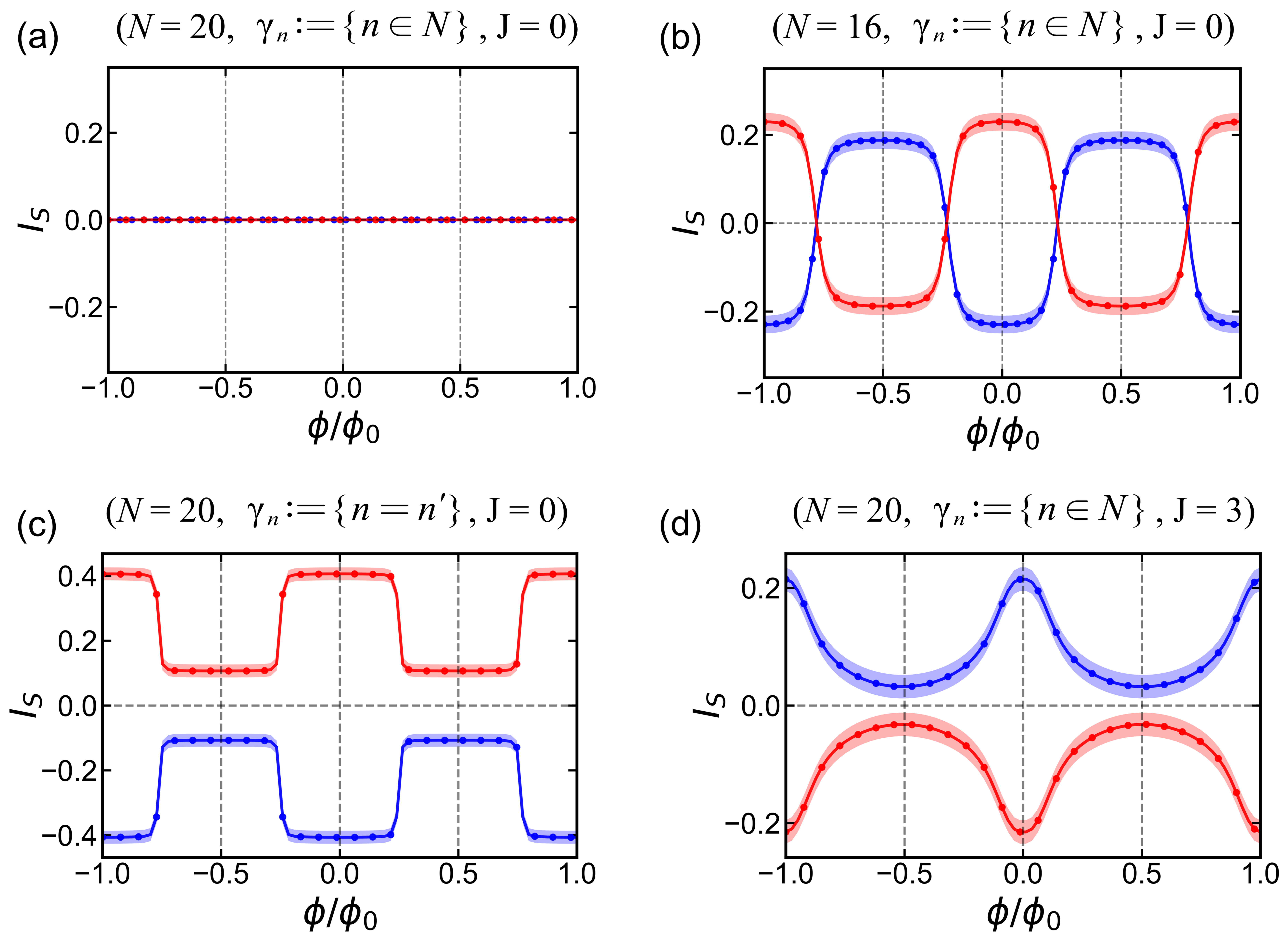}
\caption{\textcolor{black}{In the chain-internal chiral-induced SOC model, the persistent spin currents of varying chirality exhibit different behaviors under various conditions as the toroidal magnetic field strength is adjusted. Specifically, for the initial case where $N=20$, $n \in \mathcal{N}$, and $J = 0$ denoted as (a), the subsequent scenarios entail altering one of these three conditions individually to explore the changes in these physical parameters: $N=16$ (b), $n = n'$ (c), and $J = 3$ (d).}}
\label{fig4}
\end{figure}

Figure~\ref{fig4} depicts the variation of PSC with magnetic flux under different conditions, similar to what was shown in Figure~\ref{fig2}. When $\mathcal{N} \bmod M=0$, the symmetric toroidal magnetic field, and the absence of long-range hopping occur simultaneously, the PSC equals 0 regardless of whether the molecule is left-handed or right-handed (Figure~\ref{fig4} (a). When at least one of the aforementioned three conditions is not met, we obtain finite PSC values that are opposite in sign between left-handed and right-handed molecules. Initially changing the total number of lattice sites to $N=16$ while keeping the other two conditions unchanged, the resulting PSC is described by (Figure~\ref{fig4} (b)). Subsequently, in order to break the $C_2$ symmetry of the toroidal magnetic field, we replace $n$ with $n'$ in $\gamma_{x,n}$, here $n' =(5\lfloor\frac{n}{2}\rfloor+2) \delta_{1,\lfloor\frac{\bmod (n, 5)}{4}+1\rfloor \operatorname{sgn}(\bmod (n, 5))}+n \delta_{0,\lfloor\frac{\bmod (n, 5)}{4}-1\rfloor \operatorname{sgn}(\bmod (n, 5))}$, as illustrated in Figure~\ref{fig4} (c)). Lastly, we consider the inclusion of long-range hopping with $J=3$ (Figure~\ref{fig4} (d). For the model where SOC  arises from the substrate under a toroidal magnetic field, we also obtain the same results, as shown in the SI.

\begin{figure}[t]
\includegraphics[width=15cm]{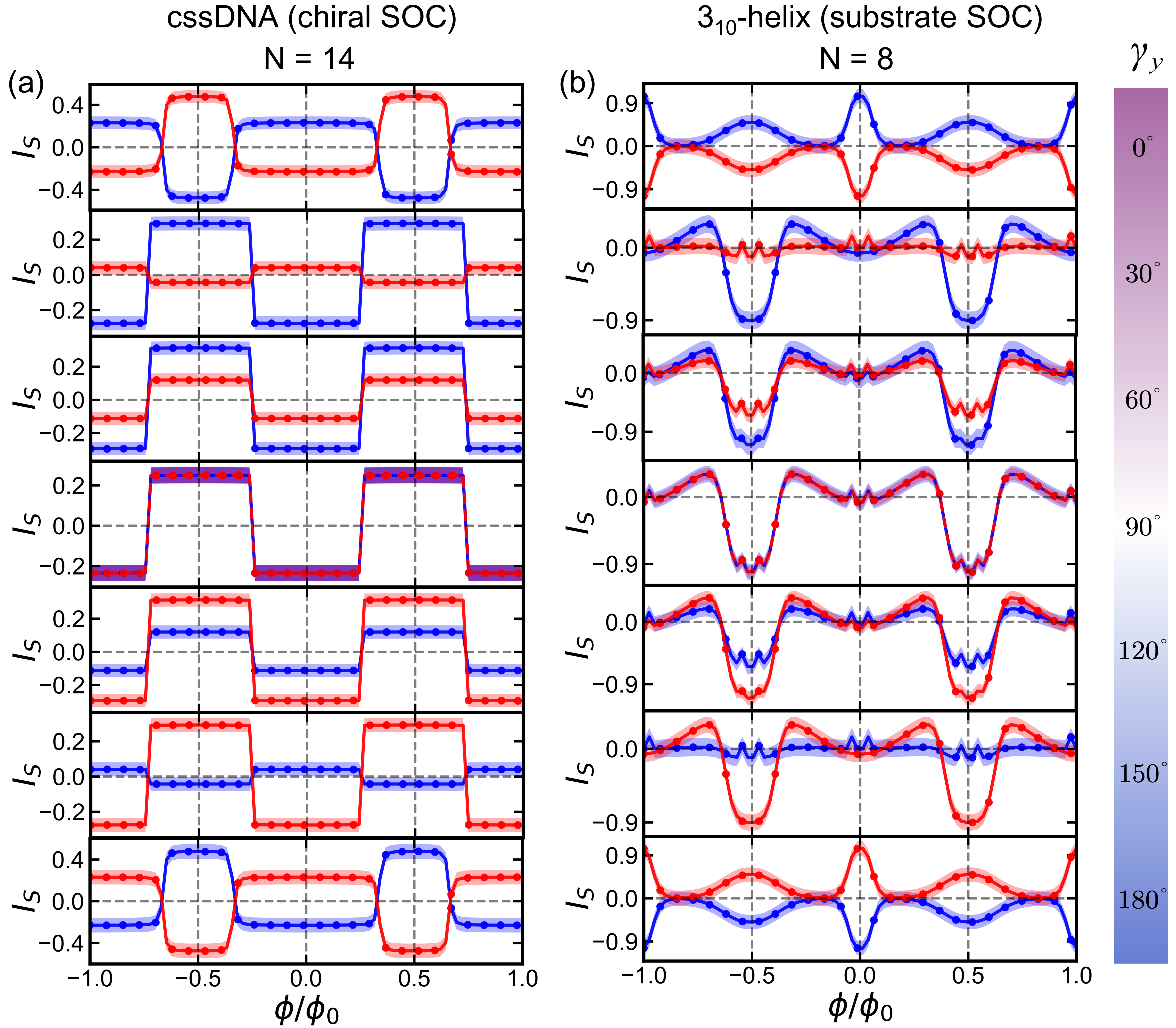}
\caption{\textcolor{black}{The variation of chiral persistent spin currents with magnetic flux for the internal molecular chirality-induced SOC model (a) and the substrate-provided SOC model (b) as the magnetic field direction relative to the molecular plane changes from 0 degrees to 180 degrees.}}
\label{fig5}
\end{figure}

\subsection{The two models under different magnetic field orientations}
Finally, we investigate the evolution of chiral PSC under various magnetic field orientations. Figure~\ref{fig5} vividly illustrates the evolution of chiral PSC in both models as the magnetic field direction relative to the chiral molecule's plane changes from 0 to 180 degrees. It's worth noting that for the specific cases of 0 and 180 degrees, we are still considering the toroidal magnetic field. Several common features can be identified in both models.  $(i)$ The PSC of left-handed and right-handed molecules with respect to the magnetic field direction $\gamma_y$ follows such a relationship: $\mathrm{I}_S^{\mathrm{left}}\left(\gamma_y\right)=\mathrm{I}_S^{\text {right }}\left(\pi-\gamma_y\right)$. $(ii)$ For $\gamma_y = 0$ or $\pi$, there is $\mathrm{I}_S^{\mathrm{left}}\left(\gamma_y\right)=-\mathrm{I}_S^{\mathrm{right}}\left(\gamma_y\right)$,  For $\gamma_y = \pi/2$, there is $\mathrm{I}_S^{\mathrm{left}}\left(\gamma_y\right)=\mathrm{I}_S^{\mathrm{right}}\left(\gamma_y\right)$, For other situations, we have $\left|\mathrm{I}_S^{\mathrm{left}}\left(\gamma_y\right)\right| \neq\left|\mathrm{I}_S^{\text {right }}\left(\gamma_y\right)\right|$. As the magnetic field direction transitions from $0$ or $\pi$ to $\pi/2$, it becomes evident that the opposite PSC under different chirality evolve towards the same value of one. In reality, the system with a magnetic field direction of $\pi/6$ can be regarded as a combination of systems with magnetic field directions of 0 and $\pi$. When the three conditions depicted in Figure~\ref{fig2} are met, the system becomes entirely equivalent to a system with a magnetic field direction of $\pi/2$ due to the absence of spin current at $\gamma_y = 0$ in this case as shown in Figure~\ref{fig4}, resulting in the emergence of trivial chiral PSC .

\subsection{Discussion on experiments}
{\textsf{Required conditions for PC in single CHMs}.} In order to observe the periodic oscillations of the PC for the CHM with an external field in experiments, it is essential to satisfy two conditions: $\phi_0 = BS$ and ensure the coherence of electrons.
Regarding the former, The strongth static magnetic field currently achievable in a laboratory setting, which is 45.22 T, readily leads to the conclusion that, for a B-DNA with a raise per residue of $3.32$$\textup{~\AA}$, at least 102 residue are needed to complete one cycle. Here, we only consider the overall magnetic flux contribution made by the ring structure when the magnetic field is perpendicular to the molecular plane. As the magnetic field direction gradually tilts relative to the molecular plane, the required number of base also increases accordingly to achieve one magnetic flux quantum. When considering a circular magnetic field of 45.22 T where the magnetic field direction is parallel to the molecular plane, meaning the magnetic flux contribution solely originates from the small loops within the molecule, a cssDNA molecule with a radius $r_0$ of 7$\textup{~\AA}$ can only reach 0.01$\phi_0$.
As for the latter, indeed, even though the coherent length of B-DNA is around 4 nm at room temperature \cite{artes}, the electronic quantum phase coherence length extends to a few hundred nanometers at low temperature \cite{kasumov}, significantly greater than the minimum circumference $l$, $l= 33.98$ nm, required to achieve one period of persistent current (PC). Consequently, we can readily identify an intermediate temperature in experiments where electron coherence is fulfilled, and the CISS effect exhibits observable behavior. Additionally, regarding the latter, we need to supplement two points, Firstly, our discussion regarding the required circumference to achieve coherence is based on the scenario of achieving one full period. However, we can also observe left and right-handed PSC within half or even smaller periods. The benefit of doing this is that, because $\phi_0 \propto l^2$, it increases the upper temperature limit, thus enhancing the visibility of the CISS effect. Secondly, although most experiments to date have shown that CISS exhibits weaker phenomena at low temperature, recent experiments have revealed some different behaviors. Yang $et$ $al.$ demonstrated that CISS polarization decreases with increasing temperature, with over $50\%$ polarization at 2 K and disappearing at 150 K \cite{1.12}. Qian $et$ $al.$, who measured the CISS effect in chiral molecule intercalation superlattices, observed over $60\%$ spin polarization at low temperatures, showing similar results \cite{1.13}. These new observed phenomena also provide us with new perspectives to achieve PC more easily in systems related to chirality.

{\textsf{The reason for neglecting the Zeeman effect}.} The main reason we are neglecting the effect of the Zeeman effect here is that it only serves to renormalize the PC values in our results and does not affect some of the conclusions we have obtained above. In the SI, we provide a more detailed explanation as well as numerical results to support this.

{\textsf{Angular persistent spin current and measurement methods for PC}.} In the preceding discussion, our primary focus has been on describing the traditional (linear) PSC that pertains to the translational motion of the spin vector. However, in reality, we also need to consider angular PSC $\mathbf{I}_\omega=\operatorname{Re} \Psi^{\dagger}(d \hat{s} / d t) \Psi$ to account for the rotational degrees of freedom of spin (precession). In the SI, we provide an analytical proof that the chiral angular PSC follows the same variation pattern as the traditional PSC with changes in the magnetic field.
For the PCC generated in quantum rings, experimental researcher mostly employ superconducting quantum interference devices or micromechanical detector to deduce PCC from the magnetic fields it generates. For aromatic or antiaromatic molecules, scientists also adopt a similar approach, using the Biot-Savart theory to infer ring charge currents from NMR spectral data. Regarding PSC, Sonin indicates that the spin torque induced by the spin current can be experimentally measured \cite{sonin}, thus enabling the detection of this equilibrium spin current. Additionally, whether it's linear or angular spin current, both can also induce electric  fields $\vec{E}_s$ and $\vec{E}_\omega$ that are experimentally detectable \cite{sun2005}, $\vec{E}_s=\frac{-\mu_0 g \mu_B}{h} \nabla \times \int \mathbf{I}_s d V \cdot \frac{\mathbf{r}}{r^3},$ and  $\vec{E}_\omega=\frac{-\mu_0 g \mu_B}{h} \int \vec{I}_\omega d V \times \frac{\mathbf{r}}{r^3}$.

{\textsf{Relationship between the persistent current and the transport current}.} When the coherent length of electrons is greater than the circumference of the CHM, the transport current within the CHM component inside the molecular junction and PC are indistinguishable in terms of both physical meaning and measurement methods, except for the difference in the driving force. For instance, they are both non-dissipative, and the charge current can induce a magnetic field, while the spin current can induce an electric field, among other similarities. Therefore, when both a magnetic field and electrodes are present, what we observe experimentally is a current within the CHM that is the sum of both contributions.

{\textsf{The influence of substrates and electrodes}.} In traditional transport experiments, a ferromagnetic substrate, such as Co or Ni, is commonly used to control the direction of spin and inject spin into the DNA layer. The heavy metal element, typically Au, within the substrate or electrode play an indispensable role in the process of measuring current, For example, in a conduction AFM experiment, one end of the tip is typically composed of heavy metals like gold or platinum, while the other end, often connected to a ferromagnetic substrate like nickel, has its top surface coated with a layer of gold thin film, approximately 5 nm thick, to prevent oxidation and thereby preserve its magnetic properties. Additionally, the inherent thiol bonds in gold enable the self-assembly of chiral molecules. Because the PC induced using a magnetic field as the driving force instead of voltage is a form of equilibrium current, the method of measuring transport current is no longer applicable here, which also allows us to eliminate the influence of external components such as electrodes and focus solely on the molecule itself. Furthermore, due to the ability to control the spin direction with a strong external magnetic field at this point, we believe that the nonmagnetic substrates, whether it's a heavy metal element like gold or a light metal like aluminum, only serves to facilitate the self-assembly of chiral molecules.

\section{Conclusions}
In summary, we develop two theoretical models with associated Hamiltonians, considering chiral and substrate-induced SOC. We investigate the CISS effect in CHM without electrodes using Green's functions. For chirality-induced SOC models, chirality-related PSC requires $\mathcal{N} \bmod M \neq 0$, asymmetric magnetic fields, or long-range hopping. This occurs with the magnetic field angle in $[0,\pi/2) \cup (\pi/2,\pi]$. For substrate-induced SOC, PCC and chirality-related PSC can form without conditions. When the magnetic field varies in $[0, \pi]$, both models show consistent transitions between these PSC types. Finally,We also discuss experimental considerations.

\section{Methods}


A more general version of the calculations for linear and angular PCC and PSC using the Green's function equation of motion can be found in Sec. A of the supplementary information for the four different models mentioned in the main text.

\begin{acknowledgement}
This work is supported by the National Natural Science Foundation of China with grant No. 11774104 and U20A2077, and partially by the National Key R\&D Program of China (2021YFC2202300). 



\end{acknowledgement}



\end{document}